\documentstyle [12pt]{article}
\begin{document}
\hyphenation{Schwarz-ian}
\hyphenation{to-po-lo-gi-cal}
\newcommand{\af}{\mbox{$\alpha$}}
\newcommand{\gh}{\mbox{$\gamma$}}
\newcommand{\de}{\mbox{$\frac{1}{2}$}}
\newcommand{\th}{\mbox{$\frac{1}{3}$}}
\newcommand{\qa}{\mbox{$\frac{1}{4}$}}
\newcommand{\sx}{\mbox{$\frac{1}{6}$}}
\newcommand{\vg}{\mbox{$\frac{1}{24}$}}

\title{On the Solution of Topological Landau-Ginzburg Models with $c=3$}
\author{\vspace{1cm} Z. Maassarani\\ \small Physics
Department\\ \small University of
Southern California\\ \small Los Angeles, CA 90089-0484}
\date{ }
\maketitle
\vspace{2cm}
\begin{abstract}
\begin{normalsize}
The solution is given for the $c=3$ topological matter model whose underlying
conformal theory has Landau-Ginzburg model $W=-\qa (x^4 +y^4)+\af x^2y^2$.
While consistency conditions are used to solve it, this model is probably
at the limit of such techniques. By using the flatness of the metric
of the space of coupling constants I rederive the differential equation
that relates the parameter \af\ to the flat coordinate $t$. This simpler method
is also applied to the $x^3+y^6$-model.
\end{normalsize}
\end{abstract}
\vspace{5cm}
\hspace{11mm}USC-91/023 \linebreak
\vspace{8mm}
\hspace{10mm}August 1991
\thispagestyle{empty}
\newpage
\setcounter{page}{1}
\section{Solution of the $x^{4}+y^{4}$-model}

I first use the methods described in \cite{dijk,nick} to compute the perturbed
topological correlation functions and their prepotential for the
superconformal field theory  that can be described by the following
Landau-Ginzburg potential:
\begin{equation}
W=-\frac{1}{4}(x^{4}+y^{4})+\alpha x^{2}y^{2}
\end{equation}
where the parameter $\alpha \;$ is complex.
When $\alpha \;$ vanishes the model reduces to the tensor product
of two minimal models with two commuting $\, U(1)$ currents.
Define $\varphi_{i}, \, i=0,...,8$ to be the chiral primary fields such
that for $\alpha \;$ vanishing we have:
$\varphi_{0}=1,\; \varphi_{1}=x,\; \varphi_{2}=y,\linebreak
\varphi_{3}=x^{2},\;
\varphi_{4}=xy,\; \varphi_{5}=y^{2},\; \varphi_{6}=x^{2}y,\;
\varphi_{7}=xy^{2},\; \varphi_8=x^{2}y^{2}$. Let $t_{i}$ be the
coupling constants corresponding to the chiral primary fields $\varphi_{i}$,
such that $t_i=0$ for $i=0$ to 8 corresponds to the unperturbed model
with $\af =0$.

The perturbed three-point functions are defined by:
\begin{equation}
C_{ijk}=\langle \varphi_{i} \varphi_{j} \varphi_{k} \exp[\sum_l t_{l} \int
d^{2}z \, G^{-}_{-\frac{1}{2}} \tilde{G}^{-}_{-\frac{1}{2}}
\varphi_{l}] \rangle \; ,
\end{equation}
{}from which one can construct a prepotential \cite{dijk,stro} $\cal F$ such
that
\begin{equation}
C_{ijk}=\frac{\partial^{3}{\cal F}}{\partial t_{i}
\partial t_{j} \partial t_{k}} \; .
\end{equation}
{}From the expansion of the exponential in (2) and the use of the symmetries of
the unperturbed potential one determines the non-vanishing
correlators. Integrating (3) one
readily gets the general form of the prepotential $\cal F$:
\begin{eqnarray}
\hspace{-6cm}{\cal F}(\vec{t}\,)\!\!\!\!&=&\de t t_0^2
+t(t_1 t_7+ t_2 t_6 + t_3 t_5) +\de t t_4^2 + t_1 t_2 t_4 f_0  \nonumber \\
  &&+\de(t_1^2 t_3 + t_2^2 t_5)f_1 +\de(t_1^2 t_5+ t_2^2 t_3)f_2
+\qa(t_1^2 t_7^2+ t_2^2 t_6^2)f_3 \nonumber \\
  &&+\qa(t_1^2 t_6^2 + t_2^2 t_7^2)f_4 +t_1 t_2 t_6 t_7 f_5
+\de (t_1 t_3^2 t_7 +t_2 t_5^2 t_6)f_6 \nonumber \\
  &&+t_3 t_5 (t_1 t_7 +t_2 t_6)f_7 +t_4(t_1 t_3 t_6 +t_2 t_5 t_7)f_8
+\de t_4^2(t_1 t_7 +t_2 t_6)f_9 \nonumber \\
  &&+\de(t_1 t_5^2 t_7+t_2 t_3^2 t_6)f_{10}+t_4(t_1 t_5 t_6
+t_2 t_3 t_7)f_{11}+\sx t_3t_5(t_3^2+t_5^2)f_{12} \nonumber \\
  &&+\vg (t_3^4+t_5^4)f_{13}+\qa t_4^2(t_3^2+t_5^2)f_{14}
+\qa t_3^2t_5^2 f_{15} \nonumber \\
  &&+\de t_3t_4^2t_5 f_{17} +\vg t_4^4 f_{19}+(higher \; order \; terms)
\end{eqnarray}
where $f_n \equiv f_n (t)$.
There are fifty four unknown functions to be determined in the
complete expression for $\cal F$ \nolinebreak !
The $f_n$'s explicitly appearing in equation (4) correspond to all the
three and four-point
functions of the relevant perturbations. The remaining $f_n$'s, which
determine the higher order perturbed functions,
can all be easily found in terms of the three and four-point functions.
The functions $f_n$ all have the
following form: \( f_n(t)=\sum_{m=0}^{\infty}a_{m}t^{2m+\overline{n}} \)
where \( \overline{n}\equiv  n \; mod \; 2 \) and \( t=t_{8} \).
These functions are determined by solving the  highly redundant set of
equations obtained from requiring that the $C_{ijk}$ be the
structure constants of an associative algebra
\begin{equation}
C_{ijp} \, g^{pq}C_{klq}=C_{ikp} \, g^{pq}C_{jlq}
\end{equation}
where \( g^{pq}=g_{pq}=C_{pq0}=\delta_{p+q,8} \) is the
$t_{i}$-independent metric on the space of the  chiral primary
fields \cite{dijk}.
The equations yielded by the conditions (5) are quadratic  in
the $f_n$ and the $f'_n$. The redundant equations serve as consistency checks.
Once the $f_n$ corresponding to three and four-point functions are
determined, the remaining functions $f_n$
can be determined by solving elementary sets of linear equations that
can be obtained from equations (5).

While solving these equations it turns out useful to introduce the ratio
\begin{equation}
\alpha (t)\equiv \frac{f_{1}(t)f_{2}(t)}{(f_{0}(t))^{2}} \; .
\end{equation}
This ratio involves the only non-vanishing three-point functions.
This apparent redefinition of $\af$ has a reason: the $\af$ of equation (6)
will turn out to be equal to that of equation (1).
After making some eliminations in a subset of equations obtained from
equations (5), one finds:
\begin{eqnarray*}
   f_5-f_7&=&\frac{\af \af'}{1-4\af^2}\; , \\
f'_7+f_7^2&=&(f_5-f_7)^2 \; ,\; \af^2(f'_5+f_5^2)=(f_5-f_7)^2 \; .
\end{eqnarray*}
These equations imply that \af\ satisfies a Schwarzian differential equation:
\begin{equation}
\{ \alpha;t \} = -(\alpha')^{2} \frac{8\alpha^{2}+6}{(1-4\alpha^{2})^{2}}
\end{equation}
where
\begin{equation}
\{ \alpha;t \} \equiv\frac{\alpha'''}{\alpha'}
-\frac{3}{2}(\frac{\alpha''}{\alpha'})^{2} \; .
\end{equation}
Taking $\alpha \;$ instead of {\em t \/} as the variable and using the
properties of
the Schwarzian derivative one can rewrite (7) as
\begin{equation}
\{ t;\alpha \} = \frac{8\alpha^{2}+6}{(1-4\alpha^{2})^{2}}\; .
\end{equation}
The general solution of equation (9) can be written as the ratio of two
independent solutions of
\begin{equation}
[(1-4 \alpha^2) \frac{d^2}{d \alpha^2} - 8\alpha \frac{d}{d \alpha}
-1]\;y=0 \; .
\end{equation}
Equation (10) can be cast into a standard hypergeometric form by
making the change of variable $z=\alpha + \frac{1}{2}$:
\begin{equation}
[z(z-1) \frac{d^2}{dz^2}+ (2z-1) \frac{d}{dz} +\frac{1}{4}]\;y =0 \; .
\end{equation}
The solutions of this equation can be expressed using the hypergeometric
function \linebreak
$F(1/2,1/2;1; \alpha +1/2)$. Equations (10) and (11) can also be derived
by the methods given in \cite{blok,saito}.
\\

A remark about the coupling constant $t$ is in order. The vanishing of
the potential
describes the target space of the SCFT. More precisely define
$\tilde{W} \equiv W+z^2$; then $\tilde{W}=0$ describes a $Z_4$ orbifoldized
torus in a
weighted projective space whose weights are (1,1,2) for
the coordinates $(x,y,z)$.
Take the modular parameter of the torus to be $\tau$. Since $\tau$ is also a
flat coordinate it must satisfy equation (9) and can therefore be written as
$\tau=\frac{a y_1+b y_2}{c y_1+d y_2}$ where $y_{1,2}$ are two independent
solutions of
equation (10) such that $t\equiv \frac{y_1}{y_2}$. One therefore has:
\begin{equation}
\tau=\frac{at+b}{ct+d} \; .
\end{equation}
To determine $(a,b,c,d)$ I utilize two facts. There is an obvious
inversion symmetry
$t \rightarrow -t$ which corresponds to $\tau \rightarrow -\frac{1}{\tau}$.
Moreover for $\af=0$ (corresponding to $t=0$) the torus has a further $Z_4$
symmetry and is
in fact rectangular; thus $t=0$ corresponds to $\tau=i$. Equation (12)
is then reduced
to equation (13)
\begin{equation}
\tau=-i\frac{\mu t+1}{\mu t-1}
\end{equation}
where $\mu$ is an undetermined scale parameter.
\\

Having found $\af(t)$, one determines:
\begin{eqnarray}
f_0&=&C(\frac{\af'}{1-4\af^2})^{1/2} \;,\; f_1=\frac{1}{2}f_0((1+2\af)^{1/2}
-(1-2\af)^{1/2}) \nonumber \\
f_2&=&\frac{1}{2}f_0 ((1+2\af)^{1/2}+(1-2\af)^{1/2}) \nonumber \\
f_3&=&2f_7=-(\frac{\af''}{\af'}+6\frac{\af \af'}{1-4\af^2}) \;,\;
f_4=\frac{1}{C^2}f_0^2     \nonumber \\
f_5&=&f_9=-\frac{1}{2}(\frac{\af''}{\af'}+4\frac{\af \af'}{1-4\af^2})
\nonumber \\
f_6&=&\frac{1}{C^2}f_2^2 \;,\; f_8=\frac{1}{C^2}f_0f_2 \;,\;
f_{10}=\frac{1}{C^2}f_1^2       \nonumber \\
f_{11}&=&\frac{1}{C^2}f_0f_1 \;,\; f_{12}=0 \;,\; f_{13}=\frac{2}{C^2}f_1f_2
\;,\; f_{14}=f_4\nonumber \\
f_{15}&=&f_3 \;,\; f_{17}=f_5 \;,\; f_{19}=-\frac{3}{2}(\frac{\af''}{\af'}
+\frac{16}{3}\frac{\af \af'}{1-4\af^2})
\end{eqnarray}
where $f_n \equiv f_{n}(t)$ and $C$ is an integration constant. The
relations $f_3=2f_7 \, , \,
f_5=f_9 \, ,..$\ and particularly
$f_{12}\equiv 0$ suggests some further symmetry of this model
that was not employed
in making the ansatz (4) for $\cal F$.
\\

To calculate the effective Landau-Ginzburg potential in the flat
coordinates $t_i \,$, one makes the most general ansatz for $W$ that
is consistent
with the symmetries of the model:
\begin{eqnarray}
\hspace{-5cm}W &=&-\qa (x^4+y^4) +\gh_0 (t) x^2 y^2  +\gh_1 (t_6 x^2 y
+t_7 x y^2)  +\gh_2 (t_3 x^2+t_5 y^2)  \nonumber \\
   &&+\gh_3 (t_5 x^2+t_3 y^2) +\de \gh_4 (t_6^2 x^2+t_7^2 y^2)
+\de \gh_5 (t_7^2 x^2+ t_6^2 y^2)           \nonumber \\
   &&+t_4 \gh_6 xy +t_6 t_7 \gh_7 xy + \gh_8 (t_1 x+t_2 y)
+\gh_9(t_3 t_7 x+t_5 t_6 y) \nonumber \\
   &&+\gh_{10}(t_5 t_7 x+t_3 t_6 y)+t_4\gh_{11}(t_6 x+t_7 y)
+\de t_6 t_7 \gh_{12}(t_6 x+t_7 y) \nonumber \\
   &&+\sx \gh_{13}(t_7^3 x+t_6^3 y)+\gh_{14}(t_1 t_7+t_2 t_6)
+t_3 t_5 \gh_{15}+\de t_4^2 \gh_{16} \nonumber \\
   &&+\de \gh_{17}(t_3^2+t_5^2)+\de \gh_{18}(t_3 t_7^2+t_5 t_6^2)
+\de \gh_{19}(t_3 t_6^2+t_5 t_7^2) \nonumber \\
   &&+t_4 t_6 t_7 \gh_{20}+\qa t_6^2 t_7^2 \gh_{21}
+\vg \gh_{22}(t_6^4+t_7^4)+t_0
\end{eqnarray}
where $\gh_i \equiv \gh_i (t)$ are, as yet, arbitrary functions.
The chiral primary fields are given by \cite{dijk}
\begin{equation}
\varphi_i (x,y)= \frac{\partial W}{\partial t_i}
\end{equation}
The structure constants $C_{ij}^{\;\;k}$ are extracted from
\begin{equation}
\varphi_i \varphi_j = C_{ij}^{\;\;k} \varphi_k \;\; mod \;\; \nabla W \; .
\end{equation}
One obtains in this way a redundant set of equations from which
one can solve for the $\gh_i$'s in terms of $\gh_0(t)$. However it is simpler
to use the values of the $C_{ijk}$ given by equations (3), (4) and
(14) to determine the $\gh_i$'s in terms of \af.
In this way one obtains additional consistency checks.
As mentionned earlier, one finds
$\gh_0 \equiv \af$. I give below the $\gh_i$ for the
terms in $W$ linear in the coupling constants $t_1$ to $t_7$:
\begin{eqnarray}
\gh_1&=&C^{-\de}(\af')^{\de}(1-4\af^2)^{\qa} \;,
\; \gh_2=\frac{1}{C}f_2 \; ,\; \gh_3=-\frac{1}{C}f_1 \nonumber \\
\gh_6&=&(\af')^{\de} \;,\; \gh_8=C^{\de}\frac{(\af')^{\de}}
{(1-4\af^2)^{\qa}} \; .
\end{eqnarray}
Given these $\gh_i$'s, solving for the remaining unknown $\gh_i$'s
is straightforward.
If one merely uses the consistency conditions on the Landau-Ginzburg
to solve for the $\gh_i$'s, one would obtain
a second possible solution for $\gh_2$ and $\gh_3 \,$:
the $\gh_2 \leftrightarrow \gh_3$ solution. However the choice
$\lim_{t_i \rightarrow 0}
\varphi_3 = x^2$ (and similarly for $\varphi_5$) made earlier, forces
$\gh_2$ to be even and $\gh_3$ to be odd in $\af$.
\\

Using the consistency conditions to solve for the $f_i$'s and
the $\gh_i$'s is extremely
tedious in general. Another procedure for obtaining the same information
can be inferred
{}from the work of \cite{nick,vafa,give}. We now give a brief exposition
of this method
and show that specific information such as the equation for $\af$
can be extracted.

\section{The flat metric method}

The basic idea in this section is to consider a general parametrization of the
Landau-Ginzburg potential, compute the metric, and impose flatness. The result
is a simpler method of obtaining the dependence of $W$ and $\cal F$ in terms of
flat coordinates. For simplicity I shall concentrate upon the behaviour of the
marginal parameter. I first study the $x^4+y^4$-model and then the
$x^3+y^6$-model.
\\

The simplest form for a Landau-Ginzburg potential is one
whose perturbation terms are linear in the coupling constants $\mu_i$,
\begin{eqnarray}
 W&=&-\qa (x^4 +y^4) + \mu_8 x^{2}y^{2} + \de \mu_7 x y^2
+ \de \mu_6 x^{2} y \nonumber \\
  & &+ \de \mu_5 y^{2}  + \mu_4 xy + \de \mu_3 x^{2}+ \mu_2 y
+ \mu_1 x + \mu_0 1
\end{eqnarray}
where $\mu_8 \equiv \af$.
The chiral primary fields  are now taken to be
$\varphi_i=\frac{\partial W(x,y;\mu)}{\partial \mu_i}$.
There is a natural metric $\tilde{g}_{ij}$ defined on the space of
coupling constants
\cite{dijk,vafa} obtained by setting $\tilde{g}_{ij}=C_{ij}^{\;\;max}$ where
$\varphi_{max}$ is the unique (up to a scaling factor) chiral primary field
of the unperturbed
theory of highest dimensional charge. The structure constants are
extracted from equations (17).
However the choice of the field $\varphi_{max}$ is ambiguous
in the following sense: any linear combination of the chiral primary
field of highest charge $(\beta \varphi_{max}
\;\rm{with}\; \beta\neq 0)$ with fields of lower charges can be used,
resulting in a conformally related metric, {\em i.e. \/}
$C_{ij}^{\;\;max}=\frac{1}{\beta}C_{ij}^{\;\;lin.\, comb.}$. However,
conformal perturbation theory gives a natural set of flat coordinates,
the $t_i$'s, in which
$g_{0\, max}=1$. This means that to get the flat metric for (19) as a function
of the flat coordinate $t$, one should take $\varphi_{max}=
(\frac{d\af}{dt}) x^2 y^2$.
For simplicity in what follows I shall take $\varphi_{max}= x^2 y^2$
and remember
to multiply the resulting metric, $\tilde{g}_{ij}\,$, by the
flattening factor \cite{give}
of $(\frac{dt}{d\af})$ later.
\\

It is easy to see that the metric elements $\tilde{g}_{ij}$ are
polynomials in the $\mu_i$'s, for $i=1$ \linebreak  to 7, with coefficient
being rational
fractions of $\mu_8=\af$. Because I am interested in the
$\mu_8$-dependence of $t$, I need only keep the linear
and quadratic terms in $\mu_1$ to $\mu_7$ as I will send these
coupling constants to zero at the end of the calculation.
The inversion of this metric is also done at the origin ($\mu_i=0$
for  $i=1$ to 7).

To obtain the differential equation for $t(\af)$ it is enough to
calculate one non-trivially vanishing component of
the Riemann tensor of the kind $R^*_{\;\, 88*}$ at the origin. The choice
$2R^4_{\;\, 884}$ simplifies the calculations by respecting the
$x\leftrightarrow y$ symmetry. The following connections
\begin{equation}
\Gamma^8_{\;\, 88} =2\Gamma^4_{\;\, 84}=\frac{t''}{t'}
\end{equation}
are the only non-vanishing terms in the  $(\Gamma)^2$-parts of
$R^4_{\;\, 884}\,$,
at the origin. The derivatives of the relevant connections are:
\begin{eqnarray}
\partial_8 \Gamma^4_{\;\, 84}&=&\de (\frac{t'''}{t'}-(\frac{t''}{t'})^2) \;,\;
\partial_4 \Gamma^4_{\;\, 88}=\frac{4\af^2+3}{(1-4\af^2)^2} \; .
\end{eqnarray}
The only contributions away from the origin come from a linear
$\mu_4$-term in $g_{48}$ and a quadratic $\mu_4$-term in $g_{88}$
which both appear in $\partial_4 \Gamma^4_{\;\, 88}$.
The vanishing of $2R^4_{\;\, 884}$ yields equation (9).
I could have chosen $R^1_{\;\, 881}$ (with $g_{17},\; g_{18},\;
g_{78}\;$ contributing) and obtained the same result.
\\

Consider now the following Landau-Ginzburg potential:
\begin{eqnarray}
W&=&-(\th x^3+\sx y^6)+ \mu_9 x y^4 +\mu_8 x y^3 +\mu_7 x y^2
+\mu_6 y^4 \nonumber \\
 & &+\mu_5 y^3 +\mu_4 xy +\mu_3 x +\mu_2 y^2 +\mu_1 y+\mu_0 1
\end{eqnarray}
where $\mu_9 \equiv \af$.
The relevant terms in the component $R^8_{\;\, 998}$ are:
\begin{eqnarray}
\Gamma^8_{\;\, 98}\! &=&\! \frac{1}{2} (32\frac{\af^2}{1-16\af^3}
+\frac{t''}{t'})\; ,\; \Gamma^9_{\;\, 99}=\frac{t''}{t'} \nonumber \\
\partial_9 \Gamma^8_{\;\, 98}\! &=&\! \frac{1}{2}
(\frac{t'''}{t'}-\frac{t''}{t'})+16\,\partial_9 (\frac{\af^2}{1-16\af^3})\; ,
\; \partial_8 \Gamma^8_{\;\, 99}=\frac{73\af+432\af^4}{(1-16\af^3)^2} \; .
\end{eqnarray}
The vanishing of $2R^8_{\;\, 998}$ yields
\begin{equation}
\{t;\af\}=2\af \frac{41-80\af^3}{(1-16\af^3)^2} \; .
\end{equation}
With the variable change $x=\frac{1}{16}\af^{-3}$ equation (24) becomes
\begin{equation}
\{t;x\}=\frac{3}{8}\frac{1}{x^2}+\frac{1}{2}\frac{1}{(x-1)^2}-\frac{31}{72}
\frac{1}{x(x-1)} \; .
\end{equation}
The solutions of equation (25) can be expressed as a ratio
of two independent solutions
of the hypergeometric equation
\begin{equation}
[x(x-1)\frac{d^2}{d x^2}+(\frac{3}{2}x-\frac{1}{2})\frac{d}{dx}
+\frac{5}{144}]\, y=0 \; .
\end{equation}
A possible choice of such independent solutions of equation (26) is: \\
$F(5/12,1/12;1/2;x) \; ,\; x^{\de}F(11/12,7/12;3/2;x)$.

\section{Conclusion}
The method of section 1 for calculating the perturbed correlators is systematic
but tedious. It becomes quickly impractical when the number of
functions $f_n$ rise sharply. Consider for instance the $x^3+y^6$-potential.
The total number of functions $f_n$ to be considered is well above 200 !!
Another obstacle arises while obtaining the consistency conditions for
the $f_n$;
equations (5) generate many redundant equations (the number of functions
$\gh_i$
is smaller but their consistency conditions have a more complicated
form). And one
often needs to consider equations involving higher order functions to get a
closed set of equations for the three and four-point functions.
One then solves for $\af(t)$, the coefficient of the marginal perturbation.
Identifying \af , for the same model, as a ratio of perturbed three-point
functions is
difficult: there are ten three-point functions and one has to consider
some four-point functions
to find the differential equation sought. In the $x^4+y^4$-model
the number of unknowns
was substantially reduced by the $x\leftrightarrow y$ symmetry but
this is not the case in general.
\\

The flat metric method is thus markedly more direct. While I have applied
the technique to determining the function $\af (t)$, one can easily determine
the other functions $\gh_i(t)$ in a general perturbation of the potential.
Using the flatness of the metric also has one other significant advantage:
the technique can be used very selectively, in that it is easy to apply it to
solving for one particular unknown in the potential.

\vspace{1cm}
\hspace{-6mm}\Large\bf Acknowledgement \\
\normalsize
\\
I would like to thank N. Warner for many enlightening discussions,
for reading the manuscript, and for advising me of some of the preliminary
results of \cite{lerc}.

\vspace{1cm}
\hspace{-6mm}\Large\bf Note Added \\
\normalsize
\\
I have been advised \cite{blok1} that there is some unpublished work by
N. Noumi in which the flat coordinates for elliptic singularities have also
been calculated.

\end{document}